\documentclass[runningheads]{llncs}
\usepackage{graphicx}
\usepackage{array}
\usepackage{subcaption}
\usepackage{pdfpages}
\usepackage{longtable}
\usepackage{booktabs}
\usepackage{multirow}
\begin{document}
\title{``It felt more real'': Investigating the User Experience of the MiWaves Personalizing JITAI Pilot Study}
\titlerunning{Investigating the User Experience of the MiWaves JITAI Pilot Study}

\author{Susobhan Ghosh\inst{1}\orcidID{0000-0003-3654-4141} \and
Pei-Yao Hung\inst{2}\orcidID{0000-0002-7415-901X} \and
Erin E. Bonar\inst{2}\orcidID{0000-0001-8849-4196} \and
Lara N. Coughlin\inst{2}\orcidID{0000-0001-6234-0850} \and
Yongyi Guo\inst{3}\orcidID{0000-0003-1192-0454} \and
Inbal Nahum-Shani\inst{2}\orcidID{0000-0001-6138-9089} \and
Maureen Walton\inst{2}\orcidID{0000-0001-6547-0204} \and
Mark W. Newman\inst{2}\orcidID{0000-0001-7186-1383} \and
Susan A. Murphy\inst{1}\orcidID{0000-0002-2032-4286}}

\authorrunning{S. Ghosh et al.}

\institute{
Harvard University, Boston, Massachusetts, USA\\
\email{susobhan\_ghosh@g.harvard.edu, samurphy@fas.harvard.edu}
\and
University of Michigan, Ann Arbor, Michigan, USA\\
\email{peiyaoh@umich.edu, erinbona@med.umich.edu, laraco@med.umich.edu, inbal@umich.edu, waltonma@med.umich.edu, mwnewman@umich.edu, peiyaoh@umich.edu}
\and
University of Wisconsin-Madison, Madison, Wisconsin, USA\\
\email{guo98@wisc.edu}
}

% \author{Paper Author}

\maketitle              % typeset the header of the contribution
\begin{abstract}
Cannabis use among emerging adults is increasing globally, posing significant health risks and creating a need for effective interventions. This paper presents an exploratory analysis of the MiWaves pilot study, a digital intervention aimed at supporting cannabis use reduction among emerging adults (ages 18-25). The findings indicate the potential of self-monitoring check-ins and trend visualizations in fostering self-awareness and promoting behavioral reflection in participants. MiWaves intervention message timing and frequency were also generally well-received by the participants. The participants' perception of effort were queried on intervention messages with different tasks, and the findings suggest that messages with tasks like exploring links and typing in responses are perceived as requiring more effort as compared to messages with tasks involving reading and acknowledging. Limitations and implications for future research in digital health and just-in-time adaptive interventions are also discussed.

\end{abstract}

\section{Introduction}

% \begin{figure*}[!t]
%     \centering
%     \includegraphics[width=0.8\textwidth]{screenshots/miwaves_overview.png}
%     \caption{Overview of the MiWaves pilot study.}
%     \label{fig:miwaves_overview}
%     \Description[Overview of the MiWaves pilot study]{Overview of the MiWaves pilot study}
% \end{figure*}

Cannabis use among emerging adults (ages 18-25) has been steadily on the rise worldwide, presenting significant health challenges \cite{unodc2023world}. 
% In the U.S., nearly 40\% of individuals in this age group reported using cannabis in 2022, reflecting a growing trend of normalization and recreational use \cite{patrick2023monitoring,SAMHSA}. 
Although not everyone who uses cannabis experiences harm, cannabis use can lead to adverse effects such as impaired cognitive function and the development of Cannabis Use Disorder (CUD) \cite{deaquino2021thc}.
% soleimani2023altered}. 
Despite these risks, there remains a significant gap in effective early interventions 
% for non-treatment-seeking individuals 
\cite{stephens2021reaching}. 
% ,lapham2019prevalence}. 
Traditional treatment approaches often fail to reach this population, 
% especially in everyday life contexts where cannabis use typically occurs. This highlights 
highlighting the need for innovative, accessible, and adaptive interventions that can provide real-time support to individuals.
% aiming to reduce their cannabis consumption.

To address this gap, prior work introduced MiWaves \cite{coughlin2024mobile}, a \emph{personalizing} Just-in-Time Adaptive Intervention (JITAI) \cite{nahum2018jitai} to help emerging adults (EAs) to self-regulate and reduce their cannabis use. MiWaves prompts participants through a smartphone app to complete self-monitoring surveys twice daily, allowing them to self-report (and in-process potentially raising their self-awareness of) their cannabis use, sleep patterns, and stress levels. Based on these self-reports, MiWaves uses a reinforcement learning (RL) algorithm to personalize the likelihood of intervention message delivery. The feasibility and acceptability of the intervention were evaluated through the MiWaves pilot study (details in Section \ref{sec:background}). 
% Figure \ref{fig:miwaves_overview} provides a visual overview of the MiWaves pilot study. 

This work contributes to the growing body of research on mobile health (mHealth) interventions for cannabis use by providing a structured analysis of participant feedback from the MiWaves pilot study.
% -- a component often overlooked in previous studies \cite{nahum2021translating,coughlin2021toward,mcclure2023feasibility,golbus2024text}. 
% In particular, we aim to analyze participant app usage behavior combined with participant feedback collected through post-test follow-up surveys, focusing on both quantitative data (e.g., engagement metrics) and qualitative insights (e.g., open-ended responses about participant experiences). \sg{TODO: List out a summary of the findings, and contributions }. Using this analysis, we hope to identify key factors that influence participant engagement and intervention effectiveness, as well as potential areas for improvement in future iterations of MiWaves.
In particular, the focus is on post-test survey data, incorporating both quantitative data (e.g., engagement metrics) and qualitative insights derived from open-ended responses around on three core themes - likes, dislikes and suggestions. 
% It is important to emphasize that this analysis was exploratory and not the primary aim of the MiWaves pilot study, which was originally designed to evaluate feasibility and acceptability. 
Although the data yielded valuable insights into participant experiences with MiWaves, 
% the data is inherently limited in richness due to the study's design that focused on feasibility and acceptability metrics. 
the richness of the data was limited by the study design, which prioritized feasibility and acceptability.
\emph{A more in-depth qualitative approach, such as interviews, could have yielded greater contextual understanding of participants' self-awareness and behavioral changes. Given these constraints of the pilot study, this exploratory analysis serves as a foundation for future work, highlighting opportunities to deepen these insights through more targeted qualitative investigations}.

The findings from the pilot study data suggest that features like self-monitoring check-ins and trend visualizations may have fostered self-awareness and prompted behavioral reflection. Participants generally found the app easy to use and integrate into daily routines, with the timing and frequency of intervention messages being well-received and viewed positively. However, intervention messages with tasks such as typing responses or exploring links, were perceived as more
% burdensome
effortful compared to the ones which required acknowledging or reading messages. Feedback highlighted opportunities to enhance future iterations by incorporating more personalized and contextually relevant message content. These insights aim to inform the design of future digital health interventions.
% In addition, they will inform the design of post-test surveys in subsequent studies, ensuring that more comprehensive and targeted qualitative data can be collected.

% \sg{TODO: Brief paper section overview}
% The paper is organized as follows: Section \ref{sec:background} provides background information on the MiWaves app and the design of the MiWaves pilot study. Section \ref{sec:related_work} reviews related work in digital health interventions and qualitative user experience analysis. Section \ref{sec:methodology} outlines the methodology of the analysis presented in this manuscript. Section \ref{sec:findings} presents the findings, focusing on self-awareness, user burden, privacy considerations, and feedback on intervention messages. Section \ref{sec:limitations_future} discusses the limitations of the study and directions for future work. Finally, Section \ref{sec:conclusion} concludes with a summary of key insights and implications for the future of MiWaves.

\section{Background}
\label{sec:background}

\begin{figure}[!ht]
    \centering
    % First subfigure
    \begin{subfigure}[b]{0.3\textwidth}
        \centering
        \includegraphics[width=\textwidth]{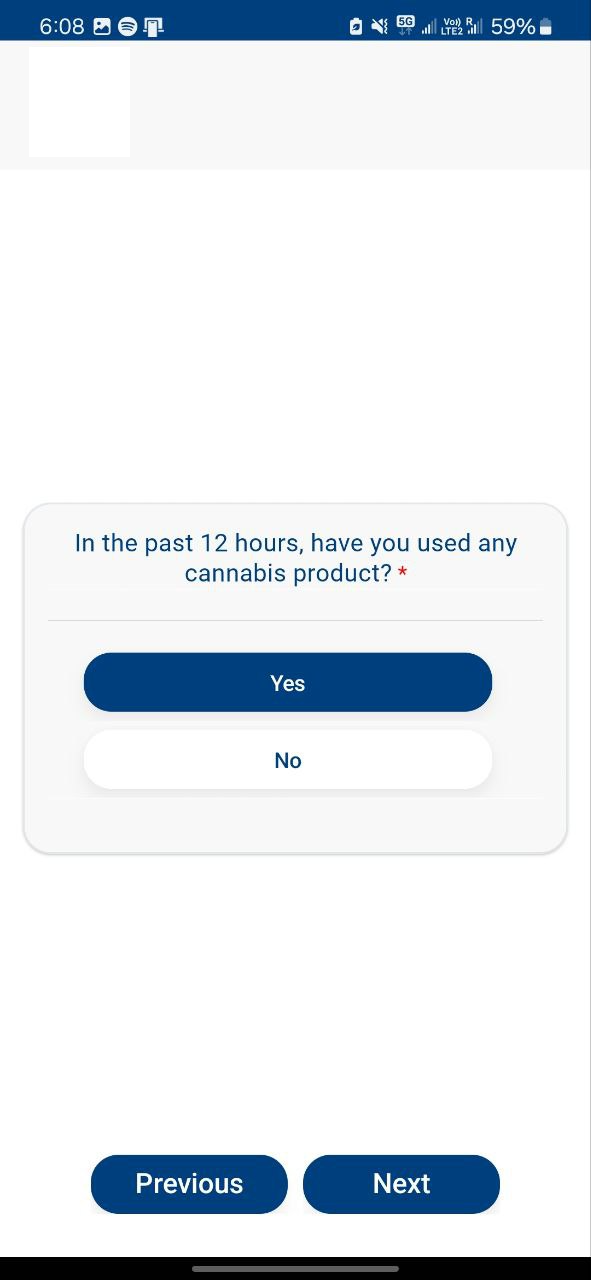} % Replace with your image file
        \caption{Self-monitoring survey}
        \label{fig:app_ema}
    \end{subfigure}
    \hfill
    % Second subfigure
    \begin{subfigure}[b]{0.3\textwidth}
        \centering
        \includegraphics[width=\textwidth]{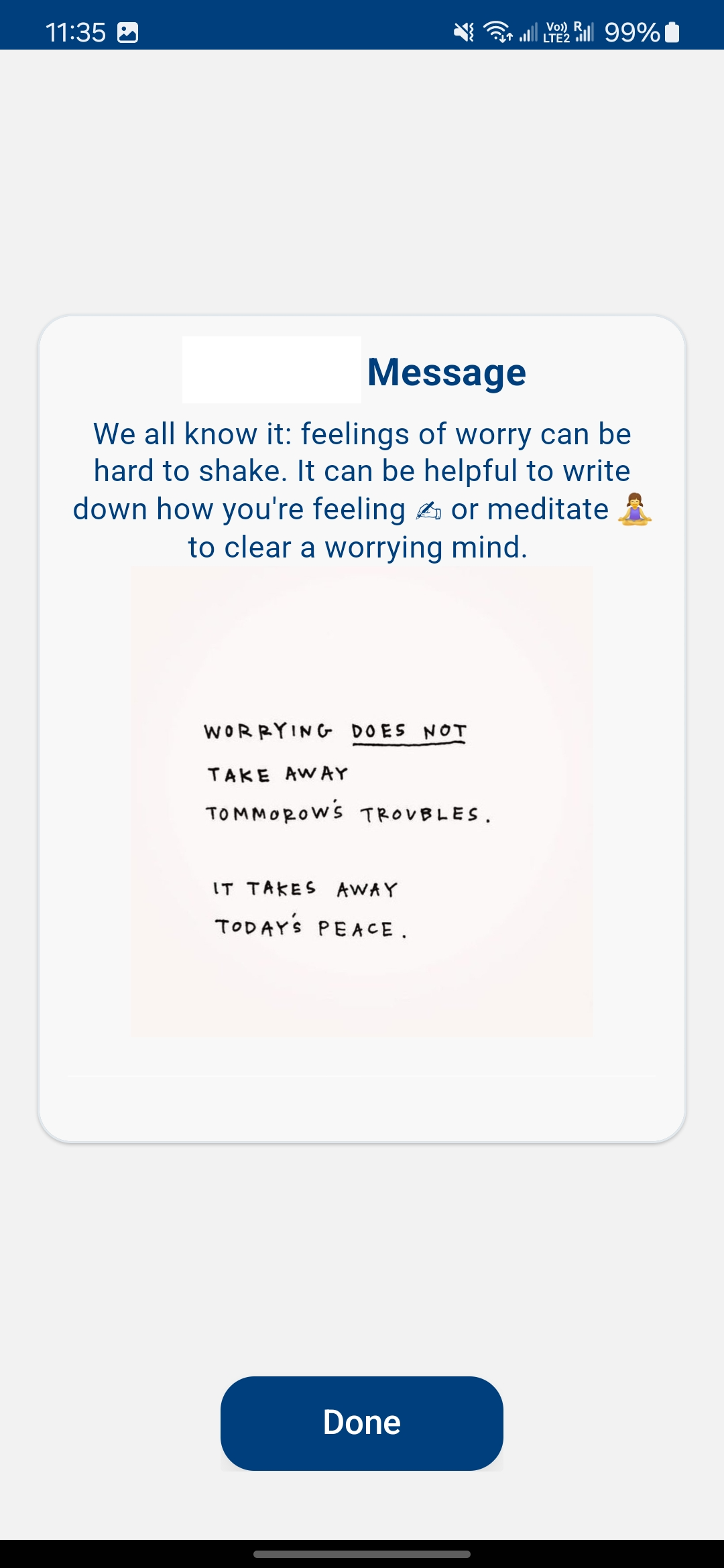} % Replace with your image file
        \caption{Intervention message}
        \label{fig:app_message}
    \end{subfigure}
    \hfill
    % Third subfigure
    \begin{subfigure}[b]{0.3\textwidth}
        \centering
        \includegraphics[width=\textwidth]{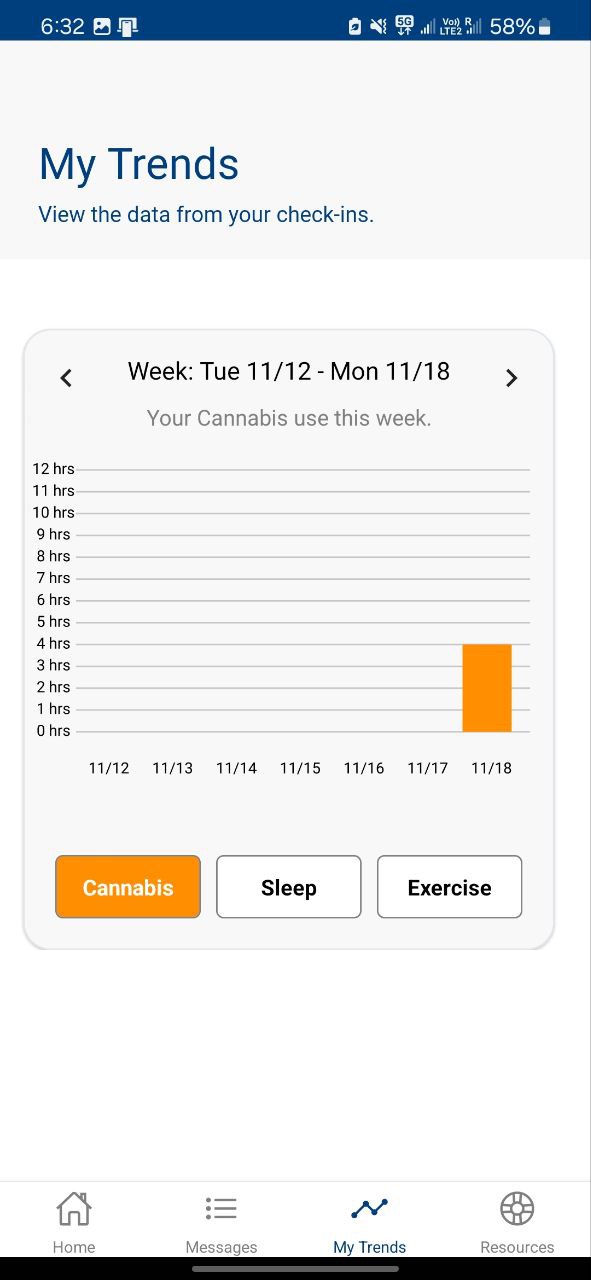} % Replace with your image file
        \caption{My Trends screen}
        \label{fig:app_life_insights}
    \end{subfigure}
    
    \caption{Screenshots from the MiWaves app. (a) Self-monitoring survey screen -- one of the questions prompting participants to report cannabis use in the past 12 hours, (b) an example of an intervention message providing motivational content to encourage reflection and mindfulness, and (c) the `My Trends' screen, which displays participants’ self-reported trends in cannabis use, sleep, and exercise over the past week, helping them reflect on their lifestyle habits.}
    \label{fig:miwaves_app_screenshots}
\end{figure}

\subsection{MiWaves app}
\label{sec:miwaves_app}
The MiWaves app (available on iOS and Android) is a digital health intervention intended to support emerging adults (EAs) in reducing cannabis use through self-monitoring, personalized feedback, and tailored resources. MiWaves utilizes \textbf{self-monitoring surveys} (see Figure \ref{fig:app_ema}), potentially prompting participants to reflect on their daily activities and experiences. During onboarding, participants can customize their preferred self-monitoring survey notification times, selecting 2-hour windows for morning and evening prompts, ensuring that the app integrates seamlessly into their daily routines. Full set of questions asked during these self-monitoring surveys can be found in Appendix C\footnote{\label{appd}https://arxiv.org/abs/2502.17645}.
% \ref{app:self_monitoring_questions}.
Based on the self-reported data, MiWaves utilizes \textbf{reBandit} \cite{ghoshmiwaves2024}, an RL algorithm, to personalize the likelihood of delivering an \textbf{intervention message} (see Figure \ref{fig:app_message}). The algorithm is designed to balance intervention delivery with moments of no intervention, maximizing user engagement while avoiding habituation to the intervention content. Since user engagement is a primary determinant of success in digital interventions \cite{nahum2018jitai}, reBandit optimizes treatment to maximize user engagement. The intervention messages are randomly chosen (without replacement) from a pool of messages varying in length and interaction type (examples provided in Table \ref{tab:intervention_prompts}).

\begin{table}[!ht]
\centering
\begin{tabular}{|>{\centering\arraybackslash}m{0.2\textwidth}|>{\centering\arraybackslash}m{0.35\textwidth}|>{\centering\arraybackslash}m{0.35\textwidth}|}
\hline
\textbf{Interaction Types} & \textbf{Short Length} & \textbf{Long Length} \\ \hline
\textbf{A (acknowledge the message)} & 
\emph{You are the artist and the future is your canvas. When you think about your life in 6 months from now, what do you hope for?} & 
\emph{Do you ever find yourself burying or suppressing your emotions? Talking your feelings out with someone you trust or writing a private journal entry can help you free those emotions.} \\ \hline
\textbf{B (participant requested to visit external resource)} & 
% \emph{What's your favorite song? Learn more about how music is beneficial to your mental health: \url{https://www.youtube.com/watch?v=zJ2YGLuzGfo}} & 
\emph{What's your favorite song? Learn more about how music is beneficial to your mental health: [youtube video link]} & 
% \emph{Trying new things doesn't have to be expensive. When you're tight on cash, check out this list to see if any of these cheap and easy hobbies seem interesting: \url{https://www.buzzfeed.com/tomvellner/cheap-easy-hobbies}} \\ \hline
\emph{Trying new things doesn't have to be expensive. When you're tight on cash, check out this list to see if any of these cheap and easy hobbies seem interesting: [weblink]} \\ \hline
\textbf{C (requires input from participant)} & 
\emph{Fun fact: even just five mins of physical activity can be beneficial. How do you get your body moving? \_\_\_\_\_\_\_\_\_\_} & 
\emph{You have the power to achieve anything you put your mind to. From this list, what are some ways you are interested in building a plan to create the future that you hope for yourself? 
(A) Writing goals 
(B) Attending therapy 
(C) Budget money 
(D) Establish healthy routines} \\ \hline
\end{tabular}
\caption{Examples of intervention messages by interaction types and length.}
\label{tab:intervention_prompts}
\end{table}

The MiWaves app also includes a \textbf{My Trends} screen (see Figure \ref{fig:app_life_insights}), which provides visual summaries of participants' self-reported cannabis use, sleep patterns, and exercise levels over the past week. This feature is intended to foster self-awareness by enabling participants to track progress and identify behavioral patterns. Additionally, MiWaves offers a comprehensive \textbf{Resources} screen, providing curated information on mental health and substance use services, overdose prevention, housing and hunger support, LGBTQ+ and gender identity resources, pregnancy and parenting services, education and employment opportunities, community activities, health services, and violence prevention resources.

\subsection{MiWaves pilot study}
\label{sec:miwaves_pilot_study}
The MiWaves pilot study was a registered clinical trial (NCT05824754) for the MiWaves intervention which ran from March 2024 to May 2024. For the study, $N=122$ EAs were recruited across the U.S. who reported using cannabis regularly and expressed motivation to reduce their use. Participants had a mean age of 21.7 years (SD = 2.0), with 53.3\% identifying as female; the racial distribution was 73.0\% White, 12.3\% Black, 4.9\% Asian, and 9.8\% Other.

This analysis primarily examines the acceptability and open-ended responses from the post-test survey administered at the end of the MiWaves pilot study. The acceptability section gathered participant feedback on the app’s functionality, aesthetics, and engagement features, as well as their perceived effort and burden, and the relevance and timing of intervention messages. The open-ended questions were intended to explore participants’ perceptions of the app, such as features they liked the most and least, and suggestions for improvement. Appendix D \footref{appd} provides the exact list of post-test survey items considered for this analysis. For the complete set of survey items or further information regarding the MiWaves pilot study design, please refer to the MiWaves protocol paper \cite{coughlin2024mobile}.

\begin{figure*}[t]
    \centering
    \includegraphics[width=0.8\textwidth]{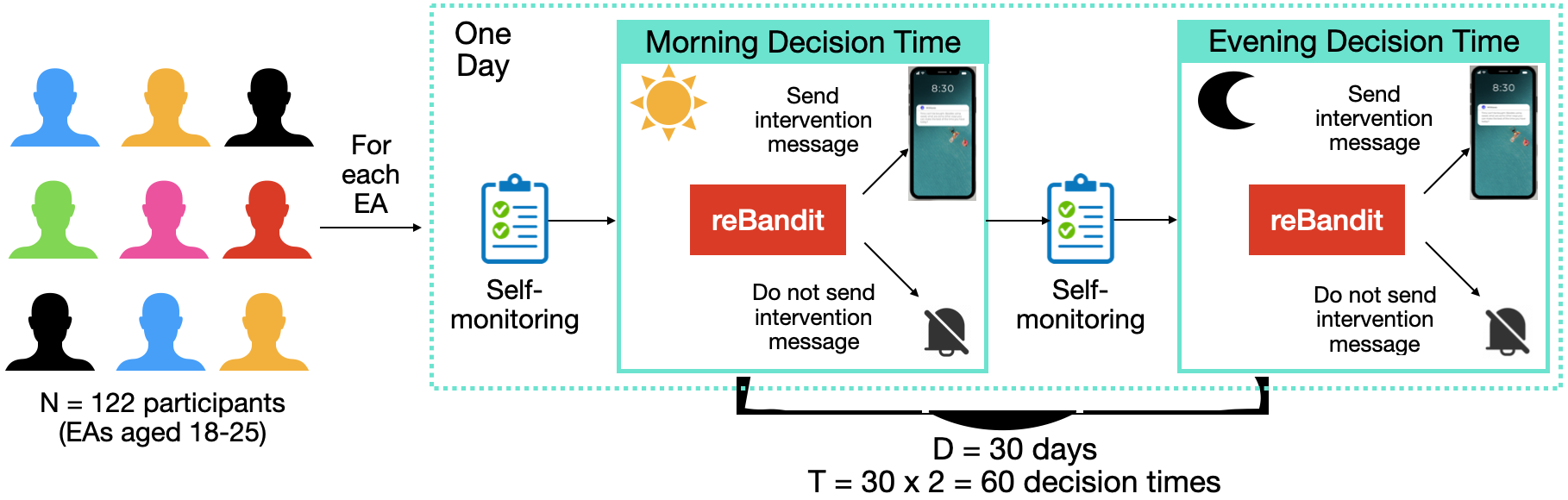}
    \caption{Overview of the MiWaves pilot study.}
    \label{fig:miwaves_overview}
\end{figure*}

\section{Related Work}
\label{sec:related_work}
This analysis builds on prior work in the space of qualitatively analyzing participant experiences with digital interventions. This aligns with broader efforts in digital health and HCI to understand how users engage with and perceive mobile interventions.

Several digital health studies have demonstrated the utility of adaptive systems for behavior change \cite{nahum2018jitai,aschentrup2024effectiveness}. 
% For instance, a meta-analysis of randomized controlled trials revealed that digital interventions effectively reduce anxiety and depression symptoms among adolescents, with treatment outcomes influenced by factors such as therapist guidance and the number of sessions \cite{li2024digital}. Additionally, research has highlighted the importance of integrating implementation science into human-centered design for digital health interventions, aiming to enhance scalability and sustained use in healthcare settings \cite{waddell2024leveraging}. 
A large volume of these works \cite{mcclure2023feasibility,klasnja2019efficacy,nahum2021translating,coughlin2021toward,golbus2024text} focus only on the quantitative metrics while reporting results. 
% For instance, HeartSteps \cite{klasnja2019efficacy,spruijt2022advancing} employed reinforcement learning (RL) to optimize the timing of interventions for increasing physical activity but largely focused on objective measures of engagement and outcomes. 
However, qualitative analysis can provide crucial insights into user experiences and identify opportunities for intervention improvement. 
% For instance, Time2Stop combined adaptive prompts with user feedback to refine its smartphone overuse intervention. The study's qualitative analysis of user feedback highlighted nuanced themes about engagement and the perceived usefulness of prompts, which informed iterative improvements to the system design \cite{time2stop2024}. Similarly,
For instance, ReVibe \cite{rabbi2019revibe} incorporated a context-assisted evening recall approach to improve self-report adherence in digital health interventions. The qualitative findings revealed how users experienced the intervention, shedding light on factors such as recall difficulties which directly influenced the design of future iterations. These studies demonstrate the critical role of qualitative analysis in uncovering insights that cannot be captured through quantitative data alone. Motivated by this line of inquiry, this present analysis applies a structured qualitative approach to participant feedback from MiWaves, with the goal of identifying themes that reflect user experience and inform opportunities for improving the intervention's design.

% In the domain of qualitative analysis of user experiences, several works have explored methodologies and their implications for improving tools and interventions. For instance, thematic analysis is commonly used to distill user feedback into actionable themes, as demonstrated in studies like User Perspectives and Ethical Experiences of Apps for Depression \cite{bowie2022user}, which analyzed reviews from app stores to uncover recurring ethical and functional issues with mental health apps. Another relevant study \cite{dennard2024systematic} synthesized qualitative research findings to identify design considerations and barriers specific to this user group. Furthermore, past research \cite{bowman2023using} has highlighted the importance of aligning qualitative methodologies with user-centered design principles to better capture the nuances of user interaction with healthcare technologies. Our work aims to utilize similar ideas to qualitatively analyze participant feedback data, while also utilizing app usage statistics to supplement our findings.

% \sg{Paragraphs need polish, sounds rough. Also cite the paper which talks about bugs being the detriment of engagement in digital interventions}

\section{Methodology}
\label{sec:methodology}
% We analyze responses from $N=112$ participants ($N=105$ filled out the open-ended questions) who completed the post-test survey, out of a total of 122 recruited participants. 
Responses from $N=112$ participants (with $N=105$ completing the the open-ended questions) were analyzed from the post-test survey administered at the conclusion of the MiWaves pilot study. These participants were drawn from a total sample of 122 individuals who participated in the pilot study.
% The MiWaves study design and protocol did not originally include qualitative analysis as a primary objective. Consequently, the qualitative analysis presented here was limited to examining the responses of the participants to open-ended questions about their likes, dislikes, and suggestions to improve the app. These questions provided valuable, albeit limited, insight into user experiences with MiWaves. 
An inductive thematic analysis \cite{clarke2017thematic} approach was applied to responses from open-ended survey items (see Appendix D.2\footref{appd}). 
Responses were initially coded into three primary domains that mirrored the structure of the survey questions: participants' likes, dislikes, and suggestions.
Within each domain, recurring patterns were identified to extract higher-level themes and actionable insights.
Analysis focused on experiences related to self-monitoring check-ins, the burden associated with intervention features, and participants' overall perceptions of usability and relevance. Sub-themes were developed to capture specific aspects of user experience, such as engagement with feedback mechanisms, the tone and timing of messages, and the perceived helpfulness of app features.
% The coded responses were classified into three primary domains (similar to the questions): participants' likes, dislikes, and suggestions. 
% Further analysis was conducted to synthesize themes and identify actionable insights. Within each domain, recurring patterns were examined, with particular attention to responses that highlighted user engagement, burden, and overall experience with the app. Subsequently, we grouped related feedback into sub-themes, such as the impact of self-monitoring check-ins, and the perception of intervention messages. 
% To deepen our understanding, we revisited the qualitative data to provide context and support for some of the quantitative survey responses. For example, participants' open-ended comments were analyzed to explain trends observed in their ratings of app usability, burden, and message relevance. This approach allowed us to draw nuanced connections between participants’ self-reported experiences and their survey responses, highlighting both strengths and areas for improvement in the MiWaves intervention. 
% These findings, along with the quantitative data analysis, informed the key insights presented in Section \ref{sec:findings}, emphasizing self-awareness, user burden, and personalization in digital health tools.
These qualitative findings were integrated with summary statistics from the quantitative survey items to inform a comprehensive interpretation of participant experience. Results are presented in Section \ref{sec:findings}, emphasizing themes of self-awareness, burden, and personalization in the context of mobile health interventions.

% In this analysis, we dived deeper into the participant dislikes, since a lot of participants mentioned encountering bugs while using the app. \sg{Bugs are a detriment to engagement in mobile heatlth studies. add cite and polish} For the quantitative analysis, we focused on responses to the acceptability questions (see Table \ref{app:quant}), which predominantly used items on the Likert scale. In addition, we correlated these responses to the participant app usage - specifically matching the responses to the interactions and dwell times of the app \sg{This part might end up going to CHI submission, won't have time for the class one}.

\section{Findings}
\label{sec:findings}
\subsection{Self-awareness: Check-ins and MyTrends}
The MiWaves app was designed to encourage self-awareness among participants by providing tools for daily self-monitoring and visualizing trends. 
% These features were central to fostering a deeper understanding of individual cannabis use patterns, and the participant feedback affirmed their effectiveness. 
Overall, participants had a 77\% engagement rate with check-ins during the study.
Notably, $N=34$ participants \emph{explicitly} mentioned that the app encouraged greater awareness of their behaviors, highlighting its potential as a behavior-change tool. It is important to note that \emph{participants were neither prompted nor asked explicitly} about changes in self-awareness during the study or in the post-test survey -- underscoring the organic emergence of this theme in open-ended responses.

% For instance, P110 remarked, \emph{``i liked how it kept me accountable with cannabis use}'', while P35 shared \emph{``daily check-ins actually forced me to confront how much weed I was smoking, since I had to ``report'' it in the morning if I did. It definitely prompted me to cut down use''}. These responses emphasize the role of various MiWaves app features in promoting self-awareness.

\subsubsection{Self-monitoring check-ins}~\ A recurring theme in participant feedback was the perceived value of the app's self-monitoring check-ins. Participants noted that these regular check-ins created a structured opportunity to think about their cannabis use and its broader implications. Specifically, $N=23$ participants highlighted the value of these check-ins in fostering self-awareness. Participant P35 expressed, \emph{``daily check-ins actually forced me to confront how much weed I was smoking, since I had to ``report'' it in the morning if I did. It definitely prompted me to cut down use.''} Similarly, P51 remarked,  \emph{``doing the check in’s made me more aware of my cannabis use, i had to recognize that i really do smoke every day and maybe it has an impact on me''}. These structured prompts served as a consistent mechanism to prompt participants to evaluate their behavior. Other participants emphasized how the check-ins fostered awareness and intentionality. For example, P72 shared, \emph{``I feel like the check-ins brought a lot of awareness to my daily habits,''} while P98 noted, \emph{``It was a daily reminder to be more intentional about my actions.''} These responses affirm that self-monitoring effectively encouraged some participants to be self-aware on their behaviors and make conscious decisions about their cannabis use.
% Several participants also appreciated the ease and accessibility of the surveys. For example, P125 shared, \emph{``I liked the phrasing and choices for the questions''}, while P116 noted, \emph{``how quick the surveys were.''} These responses indicate that the app made the process of reflection accessible and manageable for participants, to help with sustained engagement.

% Several participants appreciated how the surveys helped them explore the reasons behind their cannabis use. For example, P125 shared, \emph{``I liked the phrasing and choices for the questions,''} while P116 noted the convenience of the surveys, saying, \emph{``How quick the surveys were.''} These responses indicate that the app made the process of reflection accessible and manageable for participants, ensuring sustained engagement.

\subsubsection{My Trends}~\ The trends visualization feature also emerged as a key tool for fostering self-awareness, with $N=27$ participants specifically referencing its impact. By providing a longitudinal view of cannabis use, sleep, and stress levels, this feature may have enabled participants to recognize patterns and associations. P127 remarked that the visualization provided them \emph{``the ability to track trends over an extended period''}. Meanwhile, P130 shared \emph{``I like that we are able to track the trends of how long we used substances throughout the weeks''.} Several participants emphasized the reflective value of the visualizations. For example, P72 noted, \emph{``seeing the graphs was super cool \ldots seeing my bars for sleep lower than my bars for cannabis use definitely was eye opening''} while P8 explained, \emph{``the graphs were the most helpful thing to me because it showed the patterns in my usage.''} 
% P69 shared, \emph{``Reflecting on my substance use trends helped me see where I needed to improve,''} while P82 explained, \emph{``Tracking progress via graphs made me see where I was improving and where I needed to focus more.''} 
These responses highlight how trends visualization may have served as a powerful tool for fostering self-awareness, helping some participants better understand their behaviors and track progress over time.

% \sg{Talk about people finding trends + regular checkins to help with self-awareness}

\subsection{Burden}
User burden is a critical factor in the success of digital health interventions. High levels of burden, whether due to time-consuming tasks, rigid structures, or discomfort with sensitive questions, can lead to participant disengagement and negatively impact the effectiveness of an intervention. 
% Therefore, it is essential for interventions like MiWaves to strike a balance between collecting meaningful data and maintaining a user-friendly, low-burden experience.
% Figure \ref{fig:checkin} illustrates how 
MiWaves aimed to balance meaningful data collection with a user-friendly, low-burden experience by designing check-ins and intervention messages that participants would find both manageable and comfortable. 
% \begin{figure}[t]
%     \centering
%     \includegraphics[width=\linewidth]{figures/checkin.png}
%     \caption{Responses to checkin related questions, which asked the participant about how easy and time-consuming the self-monitoring questions (or checkins) were, and how comfortable were participants answering personal questions (eg: about their cannabis use, sleep patterns etc.)}
%     \label{fig:checkin}
%     \Description{Responses to checkin related questions, which asked the participant about how easy and time-consuming the self-monitoring questions (or checkins) were, and how comfortable were participants answering personal questions (eg: about their cannabis use, sleep patterns etc.)}
% \end{figure}
\subsubsection{Self-monitoring check-ins}
\begin{figure}[t]
    \centering
    \includegraphics[width=\linewidth]{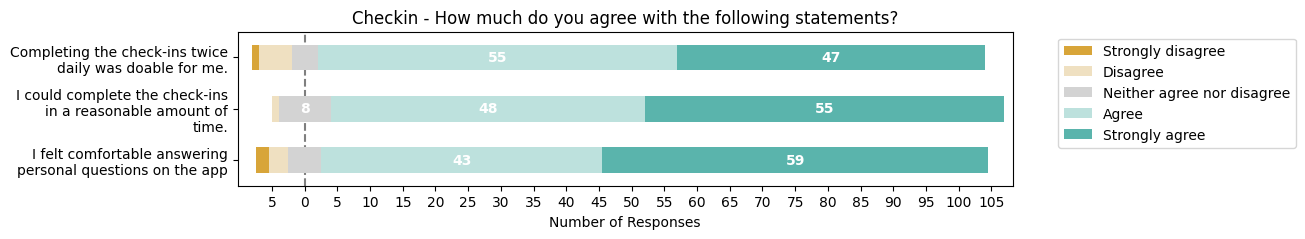}
    \caption{Responses to check-in related questions, which asked the participant about how easy and time-consuming the self-monitoring questions (or check-ins) were, and how comfortable were participants answering personal questions (eg: about their cannabis use, sleep patterns etc.)}
    \label{fig:checkin}
\end{figure}

~\ Participants rated their experience with self-monitoring check-ins on Likert-scale questions assessing perceived effort, time required, and comfort in responding to personal questions. Their response (summarized in Figure \ref{fig:checkin}) suggests that a significant majority of participants ($N=102$) reported that completing the check-ins twice daily was doable, underscoring the app’s accessibility. Similarly, the majority ($N=103$) agreed that the check-ins could be completed in a reasonable amount of time, reflecting the efficiency and quickness of the check-ins. Furthermore, participants ($N=102$) agreed that they felt comfortable answering personal questions, indicating that the check-ins were thoughtfully designed to address sensitive topics like substance use without causing undue discomfort.

Open-ended feedback closely reflected the quantitative findings related to burden and usability. Several participants emphasized the efficiency of the check-ins. For example, P116 shared, \emph{``how quick the surveys were,''}, while P135 noted that, \emph{``quick check-ins \ldots made it easy to continually use the app''}. Similarly P86 stated \emph{``I appreciated how quickly I was able to answer\ldots''}.
In addition to efficiency, participants highlighted the flexibility built into the app's design. For instance, P91 noted \emph{``I appreciate being given the grace of being able to do a check in a little late.''}, which referenced the one-hour grace period the app provided participants when they initially missed a self-monitoring check-in. 
These features likely contributed to lowering the cognitive and time burden on participants, supporting smoother integration of the app into daily routines.

\subsubsection{MiWaves Messages}
\begin{figure}[t]
    \centering
    \includegraphics[width=\linewidth]{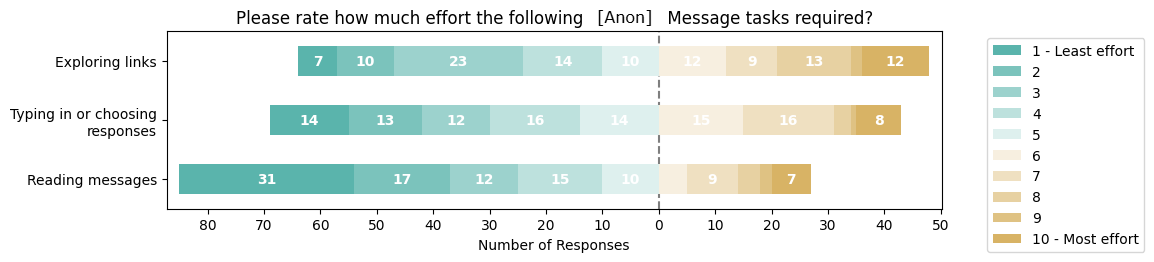}
    \caption{Responses to participant's perception of MiWaves (intervention) message burden -- specifically with respect to the different tasks in intervention messages - namely reading and acknowledging messages, visiting links and typing-in or choosing responses.}
    \label{fig:message_effort}
\end{figure}

The MiWaves app included a variety of intervention messages that prompted participants to engage through reading, exploring links, or typing in responses
% as described in Section \ref{sec:miwaves_app}
% and 
(refer Table \ref{tab:intervention_prompts}). 
% Figure \ref{fig:message_effort} highlights participants' perceptions of the effort required for these tasks. 
Participants rated their perceived effort required for these tasks on a scale of 1 (least effort) to 10 (most effort). Their response (summarized in Figure \ref{fig:message_effort}) suggests that while tasks like reading and acknowledging messages were perceived as low-effort (mean=3.79, SD=1.00), tasks such as typing-in responses (mean=4.69, SD=0.89) or exploring links (mean=5.17, SD=0.86) were viewed as more effortful. These findings suggest that, moving forward, balancing higher-effort prompts with lower-effort supportive messages may help reduce barriers to sustained participant engagement with intervention content.
% \sg{Compliment with app views for messages}.

\subsection{Personalization: MiWaves Message}
\subsubsection{Frequency and Timing}: MiWaves utilized reBandit, a reinforcement learning (RL) algorithm, to personalize the likelihood of intervention message delivery (as described in Section \ref{sec:miwaves_app}). By adapting to participant behaviors and preferences, reBandit aimed to optimize message timing to maximize engagement.
% The system contributed to a 77\% engagement rate with the check-ins during the study. 

Participants' feedback on message timing and frequency was largely positive. When asked to evaluate the number of messages received on a scale of 1 (fewer messages) to 3 (too many messages), the mean response was 1.95 (SD = 0.13), indicating that on average participants felt the frequency was appropriate and balanced. Similarly, when evaluating the convenience of message timing on a scale of 1 (not convenient) to 3 (convenient), the mean response was 2.24 (SD = 0.14), suggesting that messages were generally perceived as arriving at suitable times.
% This finding suggests that participants generally perceived the messages as arriving at convenient times.
% , though with some variability.

Overall, 72\% of participants (N = 81) reported that there were no occasions when they wanted a message but did not receive one. Similarly, 73\% (N = 82) reported that they did not receive messages at inconvenient times. Among those who did report issues, some expressed a desire for more consistent or timely messages. For instance,
% some expressed frustration when messages were not sent. For instance, 
P44 remarked, \emph{``I usually expected to receive a message after completing a survey, especially if it had been a while since my last message''} and P8 shared, \emph{``Sometimes, even when the day is going great, it's nice to have a small little note of encouragement to keep my mood up''}. These comments reflect a subset of participants who may have preferred more frequent or predictably timed messaging. However, prior research has shown that overly frequent interventions can contribute to habituation and disengagement in digital health contexts \cite{alkhaldi2016effectiveness}. A smaller number of participants noted that message timing conflicted with their daily schedules. For example, P11 stated, \emph{``Depending on work schedule, they would come during times when I was busy and could not look at the notifications in a timely manner''} and P42 mentioned, \emph{``Sometimes I would be in a meeting or in class and could not fully focus on the message!''} These comments point to opportunities for refining the algorithm to better accommodate individual availability and context.

\subsubsection{Message Content}
While MiWaves leveraged an RL algorithm to personalize the timing of intervention messages, the content of these messages was not personalized -- reflecting the pilot nature of the study. Participant feedback revealed mixed perceptions regarding the message content. Several participants described the messages being helpful and encouraging. For instanced, P105 remarked, \emph{``I liked the messages that encouraged thoughtfulness,''}, while 
% P117 shared, \emph{``They were positive and tried to be helpful''}. 
% Similarly, P122 mentioned, \emph{``I liked the positive nature of them,''} and 
P13 appreciated specific content, stating, \emph{``I liked some of the advice. The one about stress management really helped me when I was extremely stressed at work''}. Others liked the overall approach, with P48 stating, \emph{``I really liked the messages and links provided\ldots''} and P65 remarking, \emph{``The messages \ldots gave me helpful reminders when I wasn’t expecting it so it felt more real''}.

Conversely, some participants also expressed limitations in the relevance of messages.
% and message length. 
P105 stated, \emph{``Sometimes messages seemed random and not particularly helpful.''} 
% and P100 remarked, \emph{``the messages \ldots seemed unrelated''}. 
and P11 shared, \emph{``some of the messages would be unrelated to my use (ex. I don’t drink often)''}. 
% Additionally P45 stated, \emph{``The messages were tacky to be honest\ldots didn't get much at all out of them. Felt like you googled motivational phrases and just stuck em in your project''}. 
% Additionally, P118 noted, \emph{``I wish that the messages were longer and more in depth at times''}. 
Additional feedback reflected concerns about tone or emotional impact. For instance, P111 stated, \emph{``Some made me feel guilty for my habits and I really did not enjoy that''}, while 
% P101 shared that \emph{``I also thought the focus on reducing cannabis use in the messages rather than just promoting awareness of one's usage often came across as condescending''}.
P101 shared that \emph{``promoting awareness of one's usage often came across as condescending''}.

These perspectives underscore the importance of tailoring message content to individual needs and contexts in future iterations of MiWaves. Personalizing content within the framework of Just-In-Time Adaptive Interventions (JITAIs) is critical to delivering messages that are not only well-timed but also meaningful and actionable for participants. Content personalization in JITAIs remains an active area of research, with emerging approaches exploring how behavioral patterns, emotional states, and contextual signals can be integrated into adaptive messaging strategies. Leveraging such advancements may enhance user engagement and support more meaningful, sustained behavior change in digital health interventions like MiWaves.

% By adopting advancements in this area, future iterations of MiWaves can enhance engagement and intervention effectiveness, by ensuring messages resonate deeply with users while supporting long-term behavior change.

% \vspace{-1.64em}
\section{Limitations and Future Work}
\label{sec:limitations_future}

While this analysis of the MiWaves pilot study post-test data provided valuable insights, several limitations must be acknowledged. First, the pilot study was not originally designed with such analysis as the primary objective, and open-ended survey responses served as the only source of qualitative data. As a result, the depth and contextual richness of user feedback may have been limited.
% , potentially introducing recall bias
Future iterations of MiWaves may incorporate targeted questions to more systematically capture user experiences related to intervention usability, perceived impact, and engagement.
Additionally, future studies may include post-study feedback sessions with a sub-sample of participants, incorporating tailored recall aids (e.g., examples of previously received messages) to mitigate potential recall bias.
% for detailed feedback sessions post-study where participants can be shown examples of intervention messages they received during the study (to reduce recall bias). 

Another notable limitation was the participants' inability to modify self-monitoring survey times after onboarding. Introducing greater scheduling flexibility in future versions of MiWaves may help better accommodate individual preferences and routines. Technical issues, including app crashes and debugging notifications, also affected the user experience, highlighting the need for improved system stability. Lastly, while message timing was personalized using reinforcement learning, message content was not personalized, which contributed to mixed feedback. Future research could investigate content personalization strategies to enhance message relevance and engagement, ensuring interventions remain supportive rather than prescriptive.

\section{Conclusion}
\label{sec:conclusion}
% This paper presented an exploratory quantitative and qualitative analysis of participant feedback from the MiWaves pilot study, providing valuable insights into the potential of MiWaves as a digital intervention for cannabis use reduction among emerging adults. Despite study design limitations and subsequent limitations to our analysis, our initial findings highlight the promise of MiWaves as a personalized and adaptive tool for behavior change. Participants valued self-monitoring check-ins and trend visualizations for fostering self-awareness and reflection, though technical issues and limitations, such as rigid scheduling and app stability, hindered the overall experience. Privacy concerns and frustrations with expiring notifications further underscored the need for user-centric design improvements.

% To address these findings, future iterations will focus on enhancing flexibility in scheduling, addressing technical challenges, and refining engagement strategies based on user feedback. Integrating HCI-focused qualitative questions into post-intervention surveys, and leveraging post-survey memory aids will allow for deeper insights into participant experiences, while iterative advancements in personalization will ensure alignment with user needs. These improvements aim to make MiWaves a more robust, adaptive, and user-centered intervention. By incorporating lessons learned, MiWaves has the potential to advance the field of digital health interventions, setting a new standard for behavior change tools that effectively integrate into users' daily lives.
% \sg{TODO}
This paper presented an exploratory analysis of the MiWaves digital intervention, highlighting its potential as a personalized and adaptive intervention to support cannabis use reduction among emerging adults. The findings suggest that features such as self-monitoring check-ins and trend visualizations may be effective in fostering self-awareness and promoting behavioral reflection. The MiWaves intervention message timing and frequency were generally well-received. Intervention messages involving tasks such as typing responses or exploring links, were perceived as more burdensome compared to tasks like reading and acknowledging messages. 

Participant feedback revealed opportunities to enhance future interventions, particularly by enhancing the personalization and contextual relevance of intervention content. 
% Participant feedback highlighted opportunities to enhance future iterations of intervention messages by integrating more personalized and contextually relevant content and highlights
% an area for future work: understanding how message ``tone'' may
One emerging area for future research involves understanding how participants perceive and respond to variations in message ``tone'', which may influence engagement. Additional considerations include the balance between personalization and burden, and the importance of adapting content to individual preferences and contexts.
% be inferred and preferred differently across participants.
% Privacy concerns regarding possible future contextual variables and frustrations with rigid notification schedules underscore the need for user-centric design adjustments. 
% Building on these findings, future iterations of MiWaves will prioritize reducing user burden through low-effort, impactful interactions, refining the RL algorithm for intervention timing, and improving personalization. Enhanced qualitative data collection, including targeted questions and reflective post-test surveys, will help provide deeper insights into participant experiences. 
Future research may also benefit from deeper qualitative inquiry, including targeted survey items and reflective follow-up interviews, to better capture user experiences and inform design improvements in digital health interventions like MiWaves.

\section*{Acknowledgements}
Research reported in this paper was supported by NIH/NIDA 
P50DA054039, and NIH/NIBIB and OD P41EB028242. The content is solely the responsibility of the authors and does not necessarily represent the official views of the National Institutes of Health. Susan Murphy holds concurrent appointments at Harvard University and as an Amazon Scholar. This paper describes work performed at Harvard University and is not associated with Amazon. This paper grew out of a project for CS2790R - Research Topics in HCI at Harvard, and we are grateful to Prof. Katy Gero and Prof. Elena Glassman for their guidance and support.

\bibliographystyle{splncs04}
\bibliography{references}

\pagebreak
\appendix

\clearpage
\onecolumn

\section{Additional findings}
\label{app:add_findings}

% \begin{figure}[t]
%     \centering
%     \includegraphics[width=\linewidth]{figures/effort_anon.png}
%     \caption{Responses to participant's perception of MiWaves (intervention) message burden -- specifically with respect to the different tasks in intervention messages - namely reading and acknowledging messages, visiting links and typing-in or choosing responses.}
%     \label{fig:message_effort}
% \end{figure}

% \begin{figure}[t]
%     \centering
%     \includegraphics[width=\linewidth]{figures/checkin.png}
%     \caption{Responses to check-in related questions, which asked the participant about how easy and time-consuming the self-monitoring questions (or check-ins) were, and how comfortable were participants answering personal questions (eg: about their cannabis use, sleep patterns etc.)}
%     \label{fig:checkin}
% \end{figure}

\subsection{Context features and privacy}
\begin{figure}[t]
    \centering
    \includegraphics[width=\linewidth]{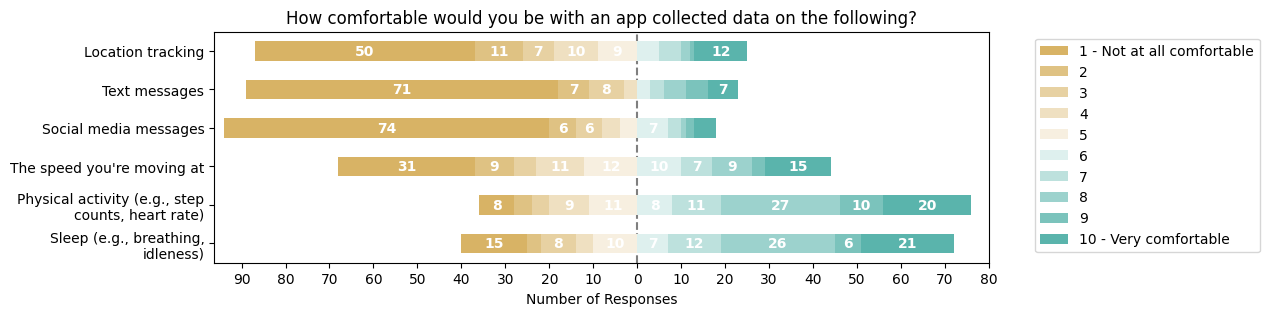}
    \caption{Responses to participant's privacy outlook - i.e. questions asking how comfortable they were with an app collecting data about their location, social media messages, text messages, moving speed, physical activity and sleep.}
    \label{fig:privacy}
\end{figure}

Digital interventions increasingly leverage contextual data, such as location, physical activity, and sleep patterns, to personalize support through adaptive algorithms. These algorithms rely on real-time data to deliver interventions tailored to users' behaviors, routines, and environments, making the experience more relevant and effective. However, integrating such data collection in interventions like MiWaves raises important considerations around privacy and user comfort which are critical for maintaining trust and engagement. 

To assess attitudes toward contextual data use, participants were asked about their comfort with an app collecting data across various contextual variables. Figure \ref{fig:privacy} presents participants' responses to questions about their comfort levels with the collection of six specific types of data - namely location tracking, text messages, social media messages, movement speed (which could include traveling in vehicles), physical activity (e.g. step count, heart rate etc.) and sleep. Participants were asked to rate their comfort level (1 to 10) with 1 being not comfortable at all, and 10 being very comfortable. The findings reveal a nuanced outlook -- the majority of participants expressed discomfort with the collection of location data (mean = 3.46, SD = 1.05), text messages (mean = 2.81, SD = 1.16), and social media messages (mean = 2.49, SD = 1.24). This suggests a substantial privacy concern among participants related to communication and geolocation data. Conversely, participants reported higher comfort levels with the collection of physical activity data (mean = 6.70, SD = 0.93) and sleep data (mean = 6.33, SD = 0.89), indicating greater acceptability of physiological and behavioral metrics. These results suggest that future implementations of adaptive digital health systems should prioritize transparency and user consent, particularly when incorporating more sensitive contextual signals. Interventions may benefit from offering customizable data-sharing options, allowing users to opt in to the types of data they are most comfortable providing.

% Interestingly, comfort levels for collecting data about speed of movement were more evenly distributed, with participants showing mixed reactions.

% \sg{Talk about future trials using a more complex context for personalization. Hence we asked participants about their outlook on which features they'd be more comfortable with }

\subsection{Expiring notifications and user experience}
In the MiWaves pilot study, the self-monitoring check-ins and intervention message notifications were designed to expire after a pre-defined window. This approach was implemented to clearly delineate time boundaries for post-study analysis, thereby improving interpretability of intervention effects. However, this design choice drew critical feedback from some participants who felt restricted, especially when they were unable to self-monitor or engage with messages.

A total of $N=8$ participants explicitly expressed frustration with the limited response windows. For instance, P102 stated, \emph{``that check ins expired too fast so i didn’t get to fill them out even though i could’ve and i wanted to''}, emphasizing the rigidity of the design. Similarly, P40 noted, \emph{``Missing the time slot to do the checkin resulted in no data for that half of a day\ldots wish there was a way to still go in and enter the data''}. Another participant, P62, expressed, \emph{``if u missed a checkin you couldn't correct your data''} reflecting broader dissatisfaction with the inability to make amendments. Suggestions for addressing this issue were also provided by participants. P102 proposed, \emph{``Longer check-in periods that don’t expire,''} while P71 suggested, \emph{``A longer gap in time to answer, 3 or 4 hours.''} Additional participants, such as P78 and P79, echoed similar sentiments, requesting \emph{``longer check-in windows''} and \emph{``longer periods that you could check in for.''} These responses indicate a clear preference for increased flexibility in response times, which could help reduce participant frustration and improve overall satisfaction.

While the expiration feature served an essential purpose for study analysis, participant feedback underscores the importance of balancing research goals with user experience. 
% For future iterations of MiWaves, one could consider allowing participants to complete expired check-ins. To maintain the rigor of effect analysis, such retroactive entries could be excluded from decision-making algorithms and flagged separately during data analysis.
Future digital health interventions might consider allowing retroactive data entry, while distinguishing these entries analytically (e.g., flagging them as post hoc) to preserve data integrity.

\section{Self-monitoring questions}
\label{app:self_monitoring_questions}

\begin{longtable}{|>{\centering\arraybackslash}m{0.2\textwidth}|>{\centering\arraybackslash}m{0.35\textwidth}|>{\centering\arraybackslash}m{0.35\textwidth}|}
% \begin{table}[!h]
\hline
\centering
% \begin{tabular}{|>{\centering\arraybackslash}m{0.2\textwidth}|>{\centering\arraybackslash}m{0.35\textwidth}|>{\centering\arraybackslash}m{0.35\textwidth}|}
% \hline
\textbf{VARIABLE NAME} & \textbf{QUESTION} & \textbf{VALUES} \\ \hline
\endfirsthead
\hline
\textbf{VARIABLE NAME} & \textbf{QUESTION} & \textbf{VALUES} \\ \hline
\endhead
use\_am (use\_pm)  & Since yesterday morning (evening), did you think about or use any suggestion from MiWaves Messages? & Y/N \\ \hline
sleep\_AM (sleep\_PM)  & Please select all hours you were asleep in the past 12 hours. & [select hours] \\ \hline
cann\_yes\_no  & In the past 12 hours, have you used any cannabis product? & Y/N \\ \hline
cann\_use\_am (cann\_use\_pm) \newline {[Display if cann\_yes\_no = Yes]} & Please select all hours you used any cannabis product in the past 12 hours. & [select hours] \\ \hline
reasons\_use \newline {[Display if cann\_yes\_no = Yes]} & What were your reasons for using cannabis? & Select all that apply \newline 1 = To enjoy the effects \newline 2 = To feel less depressed \newline 3 = To feel less anxious \newline 4 = To help sleep \newline 5 = To feel less pain \newline 6 = Nothing better to do \newline 7 = Another reason: write in \\ \hline
reasons\_not \newline {[Display if cann\_yes\_no = No]} & During the times you didn't use cannabis, what were your reasons for not using cannabis? & Select all that apply \newline 1 = Didn’t want to \newline 2 = More important things to do \newline 3 = No chance/time \newline 4 = Want to cut back \newline 5 = Ran out \newline 6 = Another reason: write in \\ \hline
drinks (only in AM)  & How many drinks containing alcohol did you have yesterday? & Type in number \\ \hline
exercise (only in AM)  & How many minutes in total did you engage in exercise yesterday? & Type in number \\ \hline
positive (only in PM) & Do you expect good things will happen to you tomorrow? & \\ \hline
% \end{tabular}
\caption{Self-monitoring questions (questions appearing in AM and PM surveys)}
% \end{table}
\end{longtable}

\begin{longtable}{|>{\centering\arraybackslash}m{0.2\textwidth}|>{\centering\arraybackslash}m{0.35\textwidth}|>{\centering\arraybackslash}m{0.35\textwidth}|}
\hline
% \textbf{VARIABLE NAME} & \textbf{QUESTION} & \textbf{VALUES} \\ \hline
% \endfirsthead
% \hline
\centering
% \begin{tabular}{|>{\centering\arraybackslash}m{0.2\textwidth}|>{\centering\arraybackslash}m{0.35\textwidth}|>{\centering\arraybackslash}m{0.35\textwidth}|}
% \hline
\textbf{VARIABLE NAME} & \textbf{QUESTION} & \textbf{VALUES} \\ \hline
% \endhead
stress  & How stressed are you right now? & 0 = Not at all \newline 1 = Slightly \newline 2 = Somewhat \newline 3 = Moderately \newline 4 = A lot \\ \hline
energy  & How energetic are you feeling? & 0 = Very low energy/Sleepy \newline 1 = Low energy/Sleepy \newline 2 = Neutral \newline 3 = High energy \newline 4 = Very energetic \\ \hline
mood  & How is your mood right now? & 0 = Very low/Negative \newline 1 = Low/Negative \newline 2 = Neutral \newline 3 = Good/Positive \newline 4 = Very good/Positive \\ \hline
social  & Right now, how would you rate your satisfaction with your social life? & 0 = Very bad/Negative \newline 1 = Bad/Negative \newline 2 = Neutral \newline 3 = Good/Positive \newline 4 = Very good/Positive \\ \hline
suds  & How anxious or distressed are you right now? & 0 = Not at all \newline 1 = Slightly \newline 2 = Somewhat \newline 3 = Moderately \newline 4 = Extremely \\ \hline
crave\_1  & Are you currently craving cannabis? & Y/N \\ \hline
crave\_2 {[Display if crave\_1 = Yes]} & Please rate your cannabis craving on the following scale: & 0 = No urge \newline 1 = Slight urge \newline 2 = Some urge \newline 3 = Moderate urge \newline 4 = Extreme urge \\ \hline
% \end{tabular}
\caption{Randomized self-monitoring questions - two questions from this pool get selected at random to appear in the self-monitoring.}
% \end{table}
\end{longtable}
% \sg{TODO}
\pagebreak

\section{Post-test survey questions}
\label{app:posttest_codebook}
\subsection{Acceptability questions (quantitative)}
\label{app:quant}

\begin{longtable}{p{6cm}ccc}
\toprule
Question & Likert Range (1-X) & Average & SD \\
\midrule
\midrule
Is the app fun to use? & 5 & 2.85 & 0.30 \\
Is the app interesting? & 5 & 2.99 & 0.30 \\
How interactive is the app? & 3 & 1.62 & 0.15 \\
\hline
How often did you have technical problems with the
app (e.g., the app crashed, content wouldn't
load)? (1 = Never, 2 = Rarely, 3 = Sometimes, 4 = Regularly) & 4 & 2.46 & 0.21 \\
\hline
Overall, how would you rate the app's appearance? & 5 & 3.01 & 0.30 \\
\hline
How true is this statement? - I felt comfortable
answering personal questions on the app (1 = Strongly disagree, 5 = Strongly agree) & 5 & 4.38 & 0.42 \\
How comfortable would you be with an app
collecting passive data such as (1 = Not at all comfortable, 10 = Very comfortable): & & & \\
Location tracking & 10 & 3.46 & 1.05 \\
Text messages & 10 & 2.81 & 1.16 \\
Social media messages & 10 & 2.49 & 1.24 \\
The speed you're moving at & 10 & 4.70 & 0.89 \\
Physical activity (e.g., step counts, heart rate) & 10 & 6.70 & 0.93 \\
Sleep (e.g., breathing, idleness) & 10 & 6.33 & 0.89 \\
\hline
How much did the ability to track your trends
increase your use of the MiWaves App? & 3 & 1.93 & 0.13 \\
How would you rate the helpfulness of your trend graphs? & 10 & 5.48 & 0.86 \\
How much did earning reward cards increase your use of the MiWaves App? & 3 & 2.87 & 0.08 \\
How would you rate the reward cards you earned? & 10 & 8.09 & 1.16 \\
\hline
Please rate how much effort the following MiWaves
Message tasks required, with 1 being least effort
and 10 being most effort: & & & \\
Exploring links & 10 & 5.17 & 0.86 \\
Typing in or choosing
responses & 10 & 4.69 & 0.89 \\
Reading messages & 10 & 3.79 & 1.00 \\
\hline
How many activities or thought exercises from
MiWaves Messages did you receive that you could
see yourself using after the end of the study? (1 = None, 5 = Most or all of them) & 5 & 2.39 & 0.33 \\
I felt the MiWaves Messages and
notifications I received in the MiWaves app showed
care, warmth, and respect (1 = Not at all, 5 = Extremely). & 5 & 3.53 & 0.32 \\
Overall, I felt that I received (1 = Fewer messages, 3 = too many messages) & 3 & 1.95 & 0.13 \\
The messages mostly came at times that were (1 = Not convenient for me, 3 = convenient for me) & 3 & 2.24 & 0.14 \\
There were specific times that I wanted to receive
a message but did not. (1 = No, 2 = Yes) & 2 & 1.28 & 0.07 \\
There were specific times that I received a
message but preferred not to. (1 = No, 2 = Yes) & 2 & 1.27 & 0.07 \\
\hline
How much do you agree with the following
statements? (1 = Strongly disagree, 5 = Strongly agree) & & &\\
Completing the check-ins twice daily
was doable for me. & 5 & 4.27 & 0.40 \\
I could complete the check-ins in a
reasonable amount of time. & 5 & 4.40 & 0.27 \\
\hline
Answer using a scale of 1 to 10, where 1 = not
at all and 10 = definitely yes: & & & \\
Would you recommend using the MiWaves app? & 10 & 7.13 & 0.80 \\
How would you rate
the helpfulness of the MiWaves app? & 10 & 6.23 & 0.89 \\
\bottomrule
\end{longtable}

\subsection{Open-ended questions}
\label{app:open_ended}
The following open-ended questions were asked to the participants:
\begin{itemize}
    \item What did you like most about the MiWaves app features and MiWaves messages?
    \item What did you like least about the MiWaves app features and MiWaves messages?
    \item What would you change about the MiWaves app?
\end{itemize}

Participants who answered `Yes' to the following questions were provided with write-in fields to describe specific instances related to their responses::
\begin{itemize}
    \item There were specific times that I wanted to receive a message but did not.
    \item There were specific times that I received a message but preferred not to.
\end{itemize}

% \section{Post-test survey questions}
% \includepdf[pages=-]{Posttest Survey Codebook_MiWaves.pdf}

\end{document}